\documentclass[aps,prl,
superscriptaddress,
showpacs,
preprintnumbers,
bibnotes,
amsmath,
amssymb,
twocolumn,
floatfix,
]{revtex4-1}

\usepackage{graphicx}% Include figure files
\usepackage{dcolumn}% Align table columns on decimal point
\usepackage{bm}% bold math
\usepackage[usenames,dvipsnames]{color}
\usepackage{verbatim}
\usepackage{amsfonts}
\usepackage{longtable}
\usepackage{natbib}
\usepackage{hyperref}% add hypertext capabilities
\usepackage[mathlines]{lineno}% Enable numbering of text and display math
\usepackage{dcolumn}
\newcolumntype{d}[1]{D{;}{.}{#1}}

\graphicspath{{figs/}}% Put all figures in a `figs' folder

\def\slashed#1{\kern+0.10em /\kern-0.50em #1}

\newcommand{\ncfigs}{216}

\begin{document}

\title{Standard Model Prediction for Direct CP Violation in $K\to\pi\pi$ Decay}

\newcommand\bnl{Physics Department, Brookhaven National Laboratory, Upton, NY 11973, USA}
\newcommand\cu{Physics Department, Columbia University, New York, NY 10027, USA}
\newcommand\pu{School of Computing \& Mathematics, Plymouth University, Plymouth PL4 8AA, UK}
\newcommand\riken{RIKEN-BNL Research Center, Brookhaven National Laboratory, Upton, NY 11973, USA}
\newcommand\edinb{SUPA, School of Physics, The University of Edinburgh, Edinburgh EH9 3JZ, UK}
\newcommand\uconn{Physics Department, University of Connecticut, Storrs, CT 06269-3046, USA}
\newcommand\soton{School of Physics and Astronomy, University of Southampton,  Southampton SO17 1BJ, UK}

\author{Z.~Bai}\affiliation{\cu}
\author{T.~Blum}\affiliation{\uconn}
\author{P.A.~Boyle}\affiliation{\edinb}
\author{N.H.~Christ}\affiliation{\cu}
\author{J.~Frison}\affiliation{\edinb}
\author{N.~Garron}\affiliation{\pu}
\author{T.~Izubuchi}\affiliation{\bnl}\affiliation{\riken}
\author{C.~Jung}\affiliation{\bnl}
\author{C.~Kelly}\affiliation{\riken}
\author{C.~Lehner}\affiliation{\bnl}
\author{R.D.~Mawhinney}\affiliation{\cu}
\author{C.T.~Sachrajda}\affiliation{\soton}
\author{A.~Soni}\affiliation{\bnl}
\author{D.~Zhang}\affiliation{\cu}

\collaboration{RBC and UKQCD Collaborations}
\noaffiliation

\date{August 11, 2015}

\begin{abstract} 
We report the first lattice QCD calculation of the complex kaon decay amplitude $A_0$ with physical kinematics, using a $32^3\times 64$ lattice volume and a single lattice spacing $a$,  with $1/a= 1.3784(68)$~GeV.  We find Re$(A_0) = 4.66(1.00)(1.26) \times 10^{-7}$ GeV and Im$(A_0) = -1.90(1.23)(1.08) \times 10^{-11}$ GeV,  where the first error is statistical and the second systematic.  The first value is in approximate agreement with the experimental result: Re$(A_0) = 3.3201(18) \times 10^{-7}$ GeV while the second can be used to compute the direct CP-violating ratio Re$(\varepsilon'/\varepsilon)=1.38(5.15)(4.59)\times10^{-4}$, which is $2.1\sigma$ below the experimental value $16.6(2.3)\times10^{-4}$.  The real part of $A_0$ is CP conserving and serves as a test of our method while the result for Re$(\varepsilon'/\varepsilon)$ provides a new test of the standard model theory of CP violation, one which can be made more accurate with increasing computer capability.
\end{abstract}


\preprint{}

 %Use showkeys class option if keyword
                              %display desired
\maketitle

The violation of CP symmetry was discovered as a subpercent admixture of the CP-even combination of $K^0$ and $\overline{K^0}$ mesons in a nominally CP-odd decay eigenstate~\cite{Christenson:1964fg}.   In the standard model this mixing is caused by a single CP-violating phase which can be introduced if there are three generations of quarks in nature~\cite{Kobayashi:1973fv}.  This CP-violating mixing is the indirect effect of virtual top quarks.  It is described by the parameter $\varepsilon$ whose measured magnitude is $2.228 (0.011) \times 10^{-3}$, a value successfully related by the standard model to the CP-violating phase measured in the decay of bottom mesons.  

Much more difficult to measure and to compute theoretically is the direct violation of CP in $K$ decay, described by the parameter $\varepsilon'$ and resulting from a CP-violating difference between the phases of the decay amplitudes $A_0$ and $A_2$, which describe kaon decay into a two-pion state with isospin $I=0$ and 2, respectively.  This direct CP violation is 3 orders of magnitude smaller than that caused by mixing, with Re$(\varepsilon'/\varepsilon) = 1.66(0.23) \times 10^{-3}$~\cite{Batley:2002gn,AlaviHarati:2002ye,PhysRevD.70.079904, Kleinknecht:2003td,Agashe:2014kda}.  Because of its small size this direct violation of CP is especially sensitive to phenomena beyond the standard model, phenomena that are believed to be required to explain the current excess of matter over antimatter in the Universe.

While standard model, direct CP violation involves massive $W$ bosons and top quarks at an energy scale far above that accessible to lattice QCD, these high-energy interactions can be accurately captured by a low-energy effective Lagrangian with Wilson coefficients ($y_i$ and $z_i$ below) which have been computed to next-leading-order in QCD and electroweak perturbation theory~\cite{Buchalla:1995vs}:
\begin{equation}
H_W = \frac{G_F}{\sqrt{2}}V_{us}^*V_{ud}\sum_{i=1}^{10} \bigl[z_i(\mu) + \tau y_i(\mu)\bigr] Q_i(\mu).
\label{eq:H_W}
\end{equation}
Here $G_F = 1.166 \times 10^{-5}/(\mathrm{GeV})^{2}$, $V_{q'q}$ is the Cabibbo-Kobayashi-Maskawa matrix element connecting the quarks $q'$ and $q$ and $\tau =  -V_{ts}^* V_{td}/V_{us}^* V_{ud}$.  The ten operators $Q_i$ are combinations of seven independent four-quark operators~\cite{Blum:2001xb}, renormalized at the scale $\mu$.  The task that remains is to compute the matrix element of the ten $Q_i$ between an initial kaon and final $\pi\pi$ state with $I=0$ or 2.  While this has been an active area for theoretical work over the past thirty years, no reliable analytic method to compute these matrix elements has emerged~\cite{Buras:1999if, Ciuchini:2000zz, Bertolini:1998vd, Pich:2004ee}.  However, this task is well suited to lattice QCD.

Over the past five years, the calculation of the $I=2$ decay has become accessible to lattice methods~\cite{Blum:2011ng,Blum:2012uk} and physical, continuum-limit results for $A_2$ are available with 10\% errors~\cite{Blum:2015ywa}.  However, calculating the $I=0$ amplitude $A_0$ faces substantial new difficulties: i) the need to create an $I=0$ two-pion state with energy well above threshold and ii) the statistical noise associated with the vacuum intermediate state.  These difficulties have been overcome by methods we will now describe.

\section{Computational Method}

The $K\to\pi\pi$ matrix elements of the ten operators $Q_i$ are determined from the Euclidean Green's functions 
\begin{equation}
C^i_{K,\pi\pi}(t_K,t_Q,t_{\pi\pi}) = \bigl\langle 0| J_{\pi\pi}(t_{\pi\pi}) Q_i(t_Q) J_K(t_K)|0 \bigr\rangle
\label{eq:3pt}
\end{equation}
in the limit of large time separations $t_{\pi\pi}-t_Q$ and $t_Q-t_K$ which projects onto the initial and final states of interest.  The operators $J_K$ and $J_{\pi\pi}$ create the initial-state kaon and destroy the two final-state pions.  Introducing a final state composed of two pions with nonzero relative momentum poses special challenges.  Using now standard methods~\cite{Lellouch:2000pv}, the desired finite-volume two-pion state would have an energy well above that with two pions at rest and require a multiexponential fit to determine the decay matrix element.  For the $I=2$, two-pion state this problem can be addressed by imposing antiperiodic boundary conditions on the down quark~\cite{Kim:2005gka,Blum:2011ng}.

However, for the $I=0$ state we must impose isospin-symmetric boundary conditions to avoid mixing the $I=0$ and 2 states.   This is possible through a major algorithmic advance: the introduction of G-parity boundary conditions~\cite{Kim:2002np, Wiese:1991ku}.  Since each pion is odd under $G$ parity, apart from the effects of their interaction, each pion must then carry a minimum momentum of $\pi/L$ for each direction (of length $L$) in which $G$ parity is imposed.  For our lattice volume, imposing $G$-parity boundary conditions in all three spatial directions results in the required $I=0$, $\pi\pi$ energy $E_{\pi\pi} \approx M_K$.

The $G$-parity transformation is described by the operator $G=Ce^{i\pi I_y}$, a product of charge conjugation $(C)$ and a $180^\circ$ isospin rotation about the $y$ axis~\cite{Lee:1956sw}.  When a lattice derivative connects quark fields across such a boundary the $(u, d)$ doublet is joined to a $G$-parity transformed doublet $(\overline{d}, -\overline{u})$.  This doubles the computational cost and requires substantial code modifications since explicit $u$ and $d$ degrees of freedom must be introduced.  In addition, the gauge fields must now obey charge-conjugation boundary conditions which demands new, special, gauge ensembles.  Since quarks and antiquarks are mixed at the boundaries, new contractions must be included in which two quark or two antiquark fields are joined by a propagator. Finally, a consistent treatment of the strange quark $s$ requires that we include an unphysical partner $s'$ to form an isodoublet that obeys $G$-parity boundary conditions~\cite{Kim:2009fe}.  When generating the $2+1$ flavor gauge ensemble we must then take the square root of the determinant of the $s-s'$ Dirac operator so that only a single strange quark flavor is included.

The second critical difficulty is that the $I=0$, two-pion state has the same quantum numbers as the vacuum, the state which thus dominates the large $t_{\pi\pi}-t_Q$ limit needed to remove excited states.  We must subtract this vacuum contribution and deal with the exponentially falling signal-to-noise ratio that results, a subtraction carried out successfully in threshold calculations, with final-state pions approximately at rest.~\cite{Blum:2011pu, QLiu_thesis}~\footnote{For a more recent, threshold calculation of $A_0$ using Wilson fermions see Ref.~\cite{Ishizuka:2015oja}}.

We reduce the noise from this vacuum subtraction using two techniques.  First, we use a split-pion operator~\cite{QLiu_thesis} to destroy the two-pion state.  Specifically, $J_{\pi\pi}(t_{\pi\pi})$ is the product of two quark-antiquark pairs, one pair at the time $t_{\pi\pi}$ and the second at $t_{\pi\pi}+4$.  By separating the pion operators we suppress 
the vacuum coupling that results when coincident pion operators immediately create and destroy a pion, reducing the vacuum noise $2\times$.  Second, we use all-to-all propagators~\cite{Foley:2005ac, Kaneko:2010ru} to construct each pion interpolating operator from a quark-antiquark pair, fixed to Coulomb gauge, with a relative coordinate, hydrogen ground-state wave function of radius $2a$ and center-of-mass coordinate distributed over a time plane at $t_{\pi\pi}$ or $t_{\pi\pi}+4$.  This choice increases the $J_{\pi\pi}$ coupling to the two-pion state relative to the vacuum, giving a further $2\times$ noise reduction~\cite{Zhang:2014ddl}.

We use a $32^3\times64$ volume, the Iwasaki$+$DSDR gauge action~\cite{Renfrew:2009wu} and M\"obius~\cite{Brower:2012vk}, domain wall fermions (DWF)~\cite{Furman:1995ky} with an extent of 12 in the fifth dimension.  By using $\beta=1.75$ and M\"obius parameters $b+c=32/12$ and $b-c=1$ we ensure that this ensemble is equivalent to our earlier dislocation-suppressing determinant ratio (DSDR) ensemble~\cite{Arthur:2012opa}, except that the latter has periodic boundary conditions and $m_\pi=170$~MeV.  Input quark masses of $m_l(=m_u=m_d)=0.0001$ and $m_s=0.045$ are used.   (If a dimensioned quantity is given without units, lattice units are implied.)  The inverse lattice spacing, residual quark mass, pion mass, and single-pion energy
are $1/a=1.3784(68)$~GeV, $m_{\mathrm{res}}=0.001842(7)$, $M_\pi=143.1(2.0)$~MeV and $E_\pi = 274.6(1.4)$~MeV.

\begin{figure}[htb]
\begin{center}
\includegraphics[width=\linewidth]{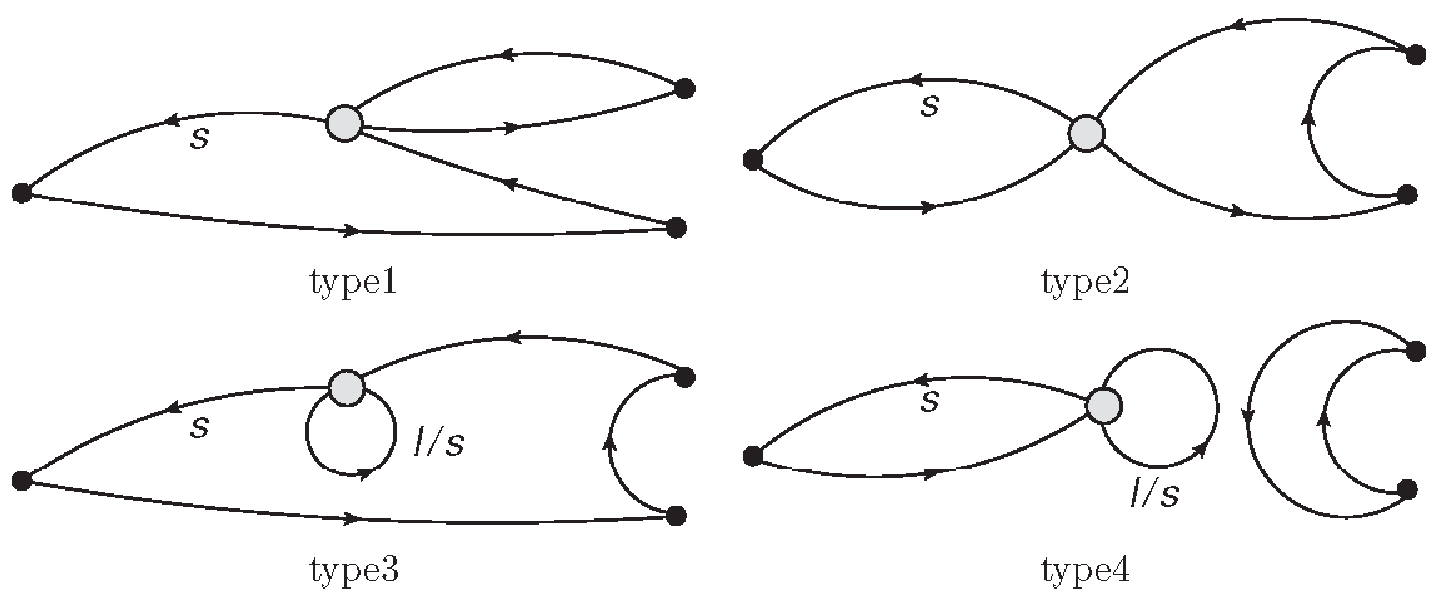}
\end{center} 
\caption{Examples of the four types of diagram contributing to the $\Delta I=1/2$, $K\to\pi\pi$ decay.  Lines labeled $\ell$ or $s$ represent light or strange quarks.  Unlabeled lines are light quarks.} 
\label{fig:diag_types}
\end{figure}

We analyzed $\ncfigs$ gauge configurations separated by four units of molecular dynamics time, starting at 300 time units for equilibration. Seventy-five distinct diagrams were computed, of four types as shown in Fig.~\ref{fig:diag_types}.  We compensated for this small number of configurations by performing 64 measurements on each configuration, introducing the kaon and pion sources on each of the 64 time planes.  (The statistically more accurate, type 1 and 2 diagrams were computed only on every eighth time plane.)  The many propagator inversions needed on each configuration were accelerated using low-mode deflation with 900 Lanczos eigenvectors~\cite{Shintani:2014vja} with the BAGEL fermion matrix package~\cite{Boyle:2009vp}.  A complete set of measurements required 20 hours on an IBM Blue Gene/Q $\frac{1}{2}$-rack~\cite{Boyle:2012iy}, in balance with the 24 hours needed to generate four time units of gauge field evolution on this same machine.

We must deal with two sorts of finite-volume effects.  The first are errors falling exponentially with increasing lattice size which result from ``squeezing'' the physical states.  Such errors are at the percent level if $L m_\pi \ge 4$.  In our case, $L m_\pi = 3.2$ and errors $\approx7\%$ may result~\cite{Blum:2012uk}.  The second are effects falling as a power of $L$, similar to the discretization of the energy that we are exploiting.  Here we apply the Lellouch-L\"uscher correction~\cite{Lellouch:2000pv} to remove the leading $1/L^3$ effect.  This requires that our final $\pi\pi$ state is an ``s-wave'' combination of the eight single-pion momenta $(\pm 1,\pm 1, \pm 1)\pi/L$.  Ensuring this s-wave symmetry requires pion operators constructed to minimize the quark-level, cubic-symmetry violations introduced by G-parity boundary conditions.

Essential to this calculation is the ability to define the seven independent, four-quark, lattice operators which correspond to those in the continuum Eq.~\eqref{eq:H_W}.  This is accomplished by using DWF whose accurate chiral symmetry ensures that the operator mixing is the same as that in the continuum.     Specifically we apply the Rome-Southampton method~\cite{Martinelli:1994ty} at $\mu = 1.53$~GeV, to introduce RI/SMOM normalization~\cite{Blum:2011pu} and then use continuum QCD perturbation theory~\cite{Lehner:2011fz}  to relate this to the Minimal Subtraction ($\overline{\mathrm{MS}}$) normalization used for the Wilson coefficients~\cite{Buchalla:1995vs}.

%and we can follow well-established procedures~\cite{Blum:2001xb, Blum:2011pu} to relate our operators to the continuum operators in Eq.~\eqref{eq:H_W}.

\section{Analysis and results}

The $K\to\pi\pi$ matrix elements of the operators $Q_i$ can be determined from the time dependence of the three-point functions defined in Eq.~\eqref{eq:3pt}:
\begin{eqnarray}
\langle J_{\pi\pi}(t_{\pi\pi})  Q_i(t_{Q})  J_K(t_K)\rangle &=&
%\nonumber \\
%&& \hskip -1.0 in = 
e^{-E_{\pi\pi}(t_{\pi\pi} - t_Q)} e^{-M_K(t_Q-t_K)} \nonumber \\
&& \hskip -1.5 in \times  \langle 0|J_{\pi\pi}(0) |\pi\pi \rangle\langle \pi\pi| Q_i(0) |K\rangle \langle K| J_K(0)|0\rangle + \cdots.
\label{eq:3point}
\end{eqnarray} 
The ellipses represent contributions from the vacuum final state or excited kaon or $\pi\pi$ states.  For the ``split-pion'' operator $J_{\pi\pi}(t_{\pi\pi})$, $t_{\pi\pi}$ is the time closest to $t_Q$.  

\begin{figure}[htb]
\begin{center}
\includegraphics[width=0.95\linewidth]{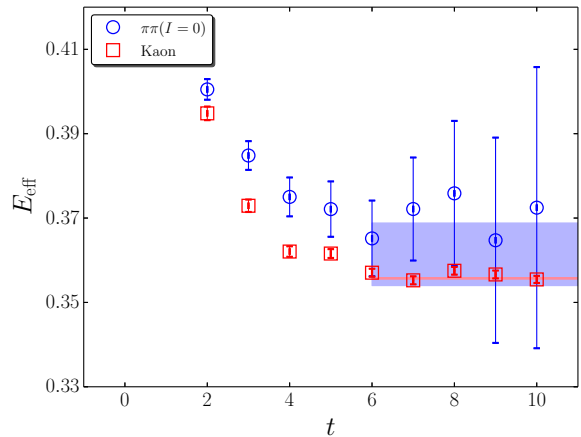}
\end{center} 
\caption{Effective energies of the kaon (squares) and two-pion (circles) states deduced from the corresponding two-point functions by equating the results from two time separations to the function $A\cosh{E_{\mathrm{eff}}(T/2-t)}$ where $T=64$ is the temporal lattice size, plotted as a function of the smallest of those two separations.  (We replace $T$ by $T-8$ for the $\pi\pi$ case.) These are overlaid by the errorbands corresponding to the fitted values of $E_{\pi\pi}$ (light blue) and $m_K$ (pink).} 
\label{fig:E_eff}
\end{figure}

The normalization factors $ \langle 0|J_{\pi\pi}(0) |\pi\pi \rangle$ and $\langle K| J_K(0)|0\rangle$ in Eq.~\eqref{eq:3point}, and the energies $M_K$ and $E_{\pi\pi}$ can be determined from the two-point functions:
\begin{equation}
\langle 0|J_X^\dagger(t_a) J_X(t_b)|0 \rangle = e^{-E_X(t_a-t_b)} \bigl|\langle 0|J_X(0)|X\rangle\bigr|^2
\label{eq:2point}
\end{equation}
where $X=\pi\pi$ or $K$.  For $X=\pi\pi$ the contribution of the vacuum intermediate state to the left-hand side must be subtracted.  Figure~\ref{fig:E_eff} shows the resulting effective energy of the kaon and two-pion states in lattice units. The kaon mass is obtained from an uncorrelated fit using $6 \le t \le 32$.  For the more challenging $I=0$, $\pi\pi$ energy, we perform a correlated, single-state fit over the interval $6 \le t \le 25$, obtaining $\chi^2/{\mathrm{dof}}=1.56(68) $.  A correlated, two-state fit using $3 \le t \le 25$ gives consistent results.  We find $M_K=490.6(2.4)$ MeV and $E_{\pi\pi}=498(11)$ MeV.  Using the L\"uscher quantization condition~\cite{Luscher:1986pf,Luscher:1990ux} we find an $I=0$, $\pi\pi$ phase shift $\delta_0 = 23.8(4.9)(1.2)^\circ$, smaller than phenomenological expectations~\cite{Colangelo:2001df,Colangelo:2015kha}.  Here the first error is statistical and the second an estimate of the $O(a^2)$ error.  For $I=2$ we find  $E_{\pi\pi}^{I=2}=573.0(2.9)$ MeV and will use $\delta_2= -11.6(2.5)(1.2)^\circ$, a corrected version of our continuum result~\cite{Blum:2015ywa}. 

Important for type 3 and 4 diagrams is the quadratically divergent quark loop.  This contribution is the same as that from the operator $\overline{d}\gamma^5 s$ with a coefficient $\propto(m_s-m_l)/a^2$.  Since $\overline{d}\gamma^5 s$ is the divergence of an axial current, its matrix element between states with equal four momentum will vanish and it will not contribute to a physical process such as $K\to\pi\pi$.  However, for matrix elements between states with unequal energies, this term may be $20\times$ larger than the other physical terms.  Even for an energy conserving amplitude, it will contribute both noise and increased systematic error from enhanced, energy nonconserving, excited-state contamination.  We determine the size of such an unphysical piece from the ratio $r_i = \langle 0|Q_i(t_Q)|K\rangle/ \langle 0|\overline{d}\gamma^5 s(t_Q)|K\rangle$ and then subtract, time slice by time slice, the operator $r_i \overline{d}\gamma^5 s(t_Q)$~\cite{Bernard:1985wf}, dramatically reducing the noise for $Q_5$, $Q_6$, $Q_7$ and $Q_8$.

\begin{figure}[htb]
\begin{center}
\includegraphics[width=0.8\linewidth]{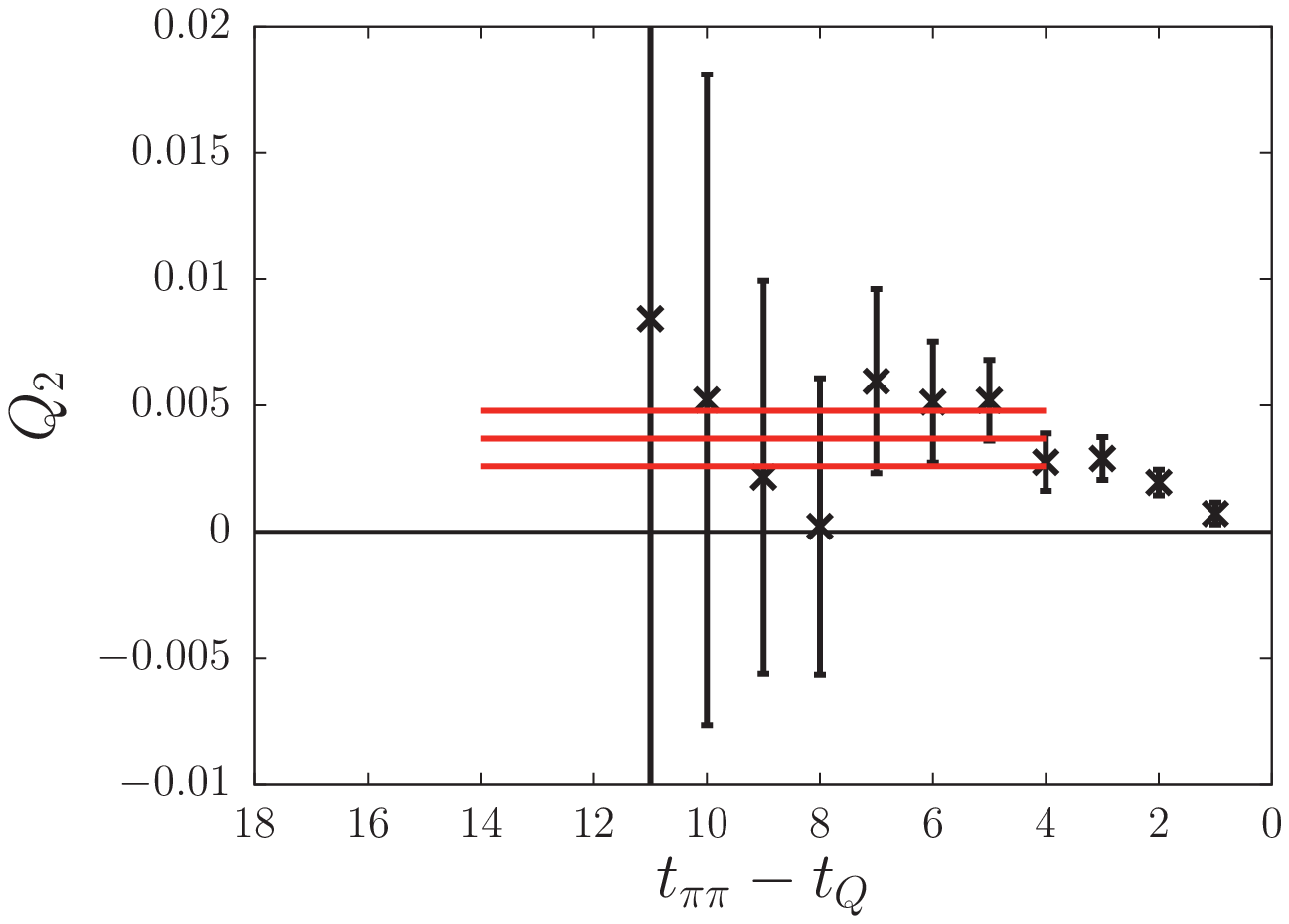}
\includegraphics[width=0.8\linewidth]{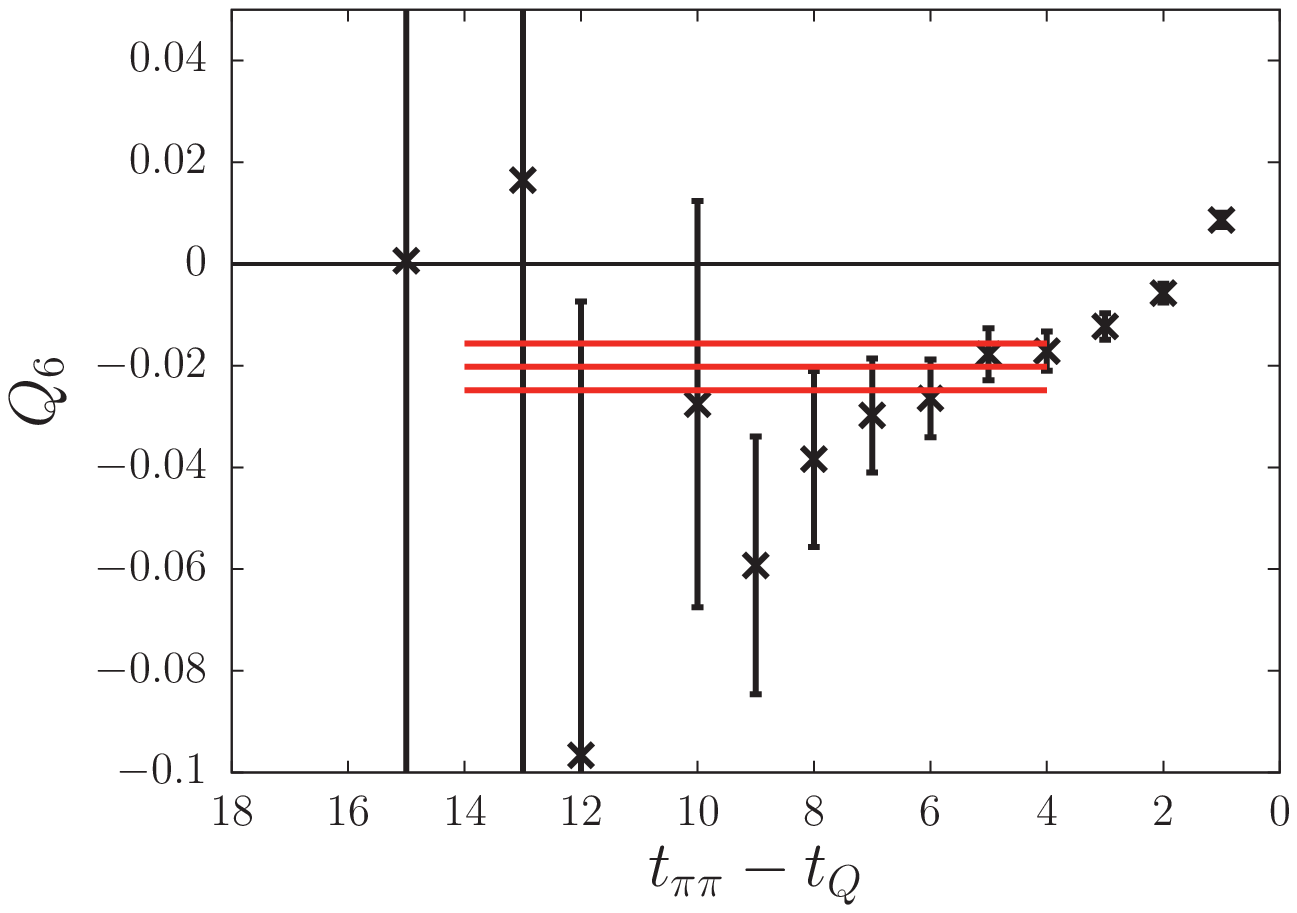}
\end{center} 
\caption{The $Q_2$ and $Q_6$ three-point functions, plotted in lattice units as functions of $t_{\pi\pi} - t_Q$, with the time dependence in Eq.~\eqref{eq:3point} removed.  The horizontal lines show the central value and errors from the fit described below.}
\label{fig:Q_eff}
\end{figure}

The largest contributions to the real and imaginary parts of $A_0$ come from $Q_2$ and $Q_6$, respectively.   Figure~\ref{fig:Q_eff} shows the three-point functions for these operators as a function of the time separation between $Q_i$ and $J_{\pi\pi}$.   Because the vacuum state may appear between these operators, the relative size of the statistical noise in the vacuum-subtracted matrix element increases rapidly as $t_{\pi\pi}-t_Q$ increases.  In Fig.~\ref{fig:Q_eff} we have combined the data (by taking an error-weighted average) from each three-point function for fixed $t_{\pi\pi}-t_Q$ and $t_Q -t_K \ge 6$.  

The matrix elements $\{\langle\pi\pi|Q_i|K\rangle\}_{1 \le i \le 10}$ are obtained by fitting the corresponding three-point functions to the time dependence in Eq.~\eqref{eq:3point}, using $t_Q-t_K \ge 6$ and $t_{\pi\pi} - t_Q \ge 4$.  We fit 25 time separations with $t_{\pi\pi} - t_K =10$, 12, 14, 16 and 18.  Figure.~\ref{fig:Q_eff} is consistent with the existence of plateaus for $t_{\pi\pi}-t_Q \ge 4$ and consistent results are obtained when including the $t_{\pi\pi} - t_Q = 3$ data, suggesting substatistical, excited-state contamination.  We estimate the systematic error from excited-state contamination as the 5\% difference between the $\pi\pi$ amplitude from a correlated, single-state fit to the $\pi\pi$ correlator with $t\ge 4$ (our matrix element fitting method) and the lowest energy amplitude found in a correlated, two-state fit to the same data with $t \ge 3$, although the difference is again within the now smaller statistical errors.  (If we omit the accurate, $t_{\pi\pi}-t_Q = 4$ data, our statistical errors increase by 40\%.)   Combining the data into bins of size 1, 2, 4 and 8 configurations, shows no bin-size dependence of the statistical errors, suggesting that autocorrelations can be neglected.  We therefore use a bin size of one.

\begin{table}[!htp]
  \centering
  \begin{tabular}{cd{22}d{20}}
    i & \multicolumn{1}{c}{Re($A_0$)(GeV)}
                                                              & \multicolumn{1}{c}{Im($A_0$)(GeV)} \\
    \hline\hline 
   1 &  1;02(0.20)(0.07)\times 10^{ -7} & 0 \\ 
    2 &  3;63(0.91)(0.28)\times 10^{ -7} & 0 \\ 
    3 & -1;19(1.58)(1.12)\times 10^{-10} &  1;54(2.04)(1.45)\times 10^{-12} \\ 
    4 & -1;86(0.63)(0.33)\times 10^{ -9} &  1;82(0.62)(0.32)\times 10^{-11} \\ 
    5 & -8;72(2.17)(1.80)\times 10^{-10} &  1;57(0.39)(0.32)\times 10^{-12} \\ 
    6 &  3;33(0.85)(0.22)\times 10^{ -9} & -3;57(0.91)(0.24)\times 10^{-11} \\ 
    7 &  2;40(0.41)(0.00)\times 10^{-11} &  8;55(1.45)(0.00)\times 10^{-14} \\ 
    8 & -1;33(0.04)(0.00)\times 10^{-10} & -1;71(0.05)(0.00)\times 10^{-12} \\ 
    9 & -7;12(1.90)(0.46)\times 10^{-12} & -2;43(0.65)(0.16)\times 10^{-12} \\ 
   10 &  7;57(2.72)(0.71)\times 10^{-12} & -4;74(1.70)(0.44)\times 10^{-13} \\ 
    \hline
  Tot &  4;66(0.96)(0.27)\times 10^{ -7} & -1;90(1.19)(0.32)\times 10^{-11} \\ 
     \hline
  \end{tabular}
  \caption{Contributions to $A_0$ from the ten continuum, $\overline{\mathrm{MS}}$ operators $Q_i(\mu)$, for $\mu=1.53$ GeV. Two statistical errors are shown: one from the lattice matrix element (left) and one from the lattice to $\overline{\mathrm{MS}}$ conversion (right).  See the Supplemental Material at [{\it URL to be inserted}] for tables of the separate matrix elements in the lattice, RI/SMOM and $\overline{\mathrm{MS}}$ schemes as well as the renormalization matrices which relate them.}
\label{tab:A0_by_op}
\end{table}

Finally these lattice matrix elements are combined with the renormalization factors, Wilson coefficients and Lellouch-L\"uscher finite-volume correction to obtain their contributions to $A_0$ as listed in  Tab.~\ref{tab:A0_by_op}.   Adding these individual contributions together gives our final result:
\begin{eqnarray}
\mathrm{Re}(A_0) &=&  4.66(1.00)(1.26) \times 10^{-7}\;\mathrm{GeV}
\label{eq:real_A0} \\
\mathrm{Im}(A_0)  &=& -1.90(1.23)(1.08) \times 10^{-11}\;\mathrm{GeV}
\label{eq:imag_A0}
\end{eqnarray}
where the first error is statistical and the second (discussed below) is systematic.  We can then compute the experimental measure of direct CP violation:
\begin{eqnarray}
\mathrm{Re}\left(\frac{\varepsilon'}{\varepsilon}\right) &=& \mathrm{Re}\left\{ \frac{i\omega e^{i(\delta_2-\delta_0)}}{\sqrt{2}\varepsilon}\left[ \frac{\mathrm{Im}A_2}{\mathrm{ReA}_2}-\frac{\mathrm{Im}A_0}{\mathrm{Re}A_0} \right] \right\}
\label{eq:ep-e_theory} \\
&=&1.38(5.15)(4.59)\times10^{-4},
\label{eq:ep-e}
\end{eqnarray}
obtained using the Im$(A_0)$  and $\delta_0$ values given above and our earlier results for Im$(A_2)$ and $\delta_2$~\cite{Blum:2015ywa}.  We use the experimental values for Re$(A_0)$, Re$(A_2)$ and their ratio $\omega$ (since these are accurately determined from the measured $K\to\pi\pi$ decay rates) and the experimental value for $\varepsilon$.

We now briefly describe the systematic error estimates given in Tab.~\ref{tab:sys_errors}; more complete explanations will appear in a later paper.  We estimate the finite lattice spacing error by averaging the differences between the three, individual $\Delta I = 3/2$, $K\to\pi\pi$  matrix elements obtained using the present gauge action~\cite{Blum:2012uk} and our recent continuum-limit results~\cite{Blum:2015ywa}.  The errors arising from the Wilson coefficients are estimated as the difference of our result computed using the leading-order (LO) and next-to-leading-order (NLO) formulae for Re$(A_0)$~\cite{Buchalla:1995vs}.   A similar uncertainty arises when we relate our lattice operators to the $\overline{\mathrm{MS}}$ operators in the continuum expression for $H_W$.  This procedure is compromised by our use of NLO perturbation theory at $\mu=1.53$~GeV to relate the RI- and $\overline{\mathrm{MS}}$-normalized operators and by our omission of dimension-5 and 6 quark-bilinear operators (whose contribution we expect to be small) from the nonperturbative operator matching.  These operator normalization errors are estimated, as in Ref.~\cite{Blum:2015ywa}, by comparing two different RI/SMOM schemes.  Parametric uncertainties are found by propagating the standard model input parameter errors.  Comparing two ans\"atze for the $E_{\pi\pi}$ dependence of $\delta_0$ suggests a 11\% uncertainty in the Lellouch-L\"uscher finite-volume correction.  Finally systematic errors are introduced by our mildly unphysical kinematics which are estimated from a companion calculation using a 10\% larger value of the strange quark mass.

\begin{table}[!htp]
  \centering
  \begin{tabular}{lr|lr}
   Description & Error & Description & Error\\
    \hline\hline
Finite lattice spacing		& 12\% 		& Finite volume 			&   7\% \\
Wilson coefficients			& 12\%		& Excited states			&   $\le5$\% \\
Parametric errors			&  5\% 		& Operator renormalization	& 15\% \\ 
Unphysical kinematics            &   $\le3$\% 	& Lellouch-L\"uscher factor	& 11\% \\
    \hline
\multicolumn{3}{l}{Total (added in quadrature)}						& 27\% \\
    \hline
  \end{tabular}
  \caption{Representative, fractional systematic errors for the individual operator contributions to Re$(A_0)$ and Im$(A_0)$.}
\label{tab:sys_errors}
\end{table}

\section{Conclusion}

We have presented the first calculation of the direct CP violation parameter $\varepsilon'$ with controlled errors.    While the $2.1\sigma$ difference between our value for Re$(\varepsilon'/\varepsilon)$ and experiment gives a strong motivation to refine the present calculation, we believe that the absolute size of our statistical and systematic errors demonstrates that this is now a quantity accessible to lattice QCD.  Also for the first time, we have computed the real part of the decay amplitude $A_0$.  The result agrees with the experimental value and provides a test of our methods.   This result for Re$(A_0)$ is  consistent with our earlier explanation of the $\Delta I=1/2$ rule~\cite{Boyle:2012ys} in which the large ratio of Re$(A_0)$/Re$(A_2)$ resulted from a significant cancellation between the two dominant terms contributing to Re$(A_2)$, a cancellation which does not occur for Re$(A_0)$.  We emphasize that this calculation can be substantially improved by adding more statistics and by studying larger volumes and additional lattice spacings to better control the large systematic errors.  Nonperturbative, step-scaling methods can relate the lattice operators being used to those defined at much smaller lattice spacing where the perturbative Wilson coefficients can be more accurately determined.  We expect that a 10\% error relative to the measured value of Re$(\varepsilon'/\varepsilon)$ can be achieved within 5 years, motivating continued improvement in the experimental result.  Substantially more accurate results will become possible with further increases in computer power and the inclusion of electromagnetism. 

\section{Acknowledgments}

We would like to thank our RBC and UKQCD collaborators for helpful discussions and support.  This calculation was carried out under the INCITE Program of the US DOE on the IBM Blue Gene/Q (BG/Q) Mira machine at the Argonne Leadership Class Facility, a DOE Office of Science Facility supported under Contract De-AC02-06CH11357, on the STFC-funded ``DiRAC'' BG/Q system in the Advanced Computing Facility at the University of Edinburgh, on the BG/Q computers of the RIKEN BNL Research Center and the Brookhaven National Laboratory.  The DiRAC equipment was funded by BIS National e-infrastructure capital grants ST/K000411/1, STFC capital grant ST/H008845/1, and STFC DiRAC Operations grants ST/K005804/1 and ST/K005790/1.  DiRAC is part of the National e-Infrastructure.  Z.B., N.H.C., R.D.M.~and D.Z.~are supported in part by U.S. DOE grant \#De-SC0011941,  P.A.B.~and J.F.~from the STFC grants ST/L000458/1 and ST/J000329/1 and T.B.~by the US Department of Energy Grant No. De-FG02-92ER41989.   T.I., C.J., C.L. and A.S.~are supported in part by US DOE Contract \#AC-02-98CH10886(BNL) while T.I is also supported by Grants-in-Aid for Scientific Research \#26400261.  C.K.~is supported by a RIKEN foreign postdoctoral research (FPR) grant, N.G by Leverhulme Research grant RPG-2014-118 and C.S.~was partially supported by UK STFC Grants ST/G000557/1 and  ST/L000296/1.

\bibliography{refs}

\clearpage
\newpage

\setcounter{page}{1}
\renewcommand{\thepage}{Supplementary Information -- S\arabic{page}}
\setcounter{table}{0}
\renewcommand{\thetable}{S\,\Roman{table}}
\setcounter{equation}{0}
\renewcommand{\theequation}{S\,\arabic{equation}}

In this section we provide additional technical information regarding our results.  A complete account of the calculation will be published in a separate paper.

\subsection{Lattice matrix elements}

\begin{table}
\begin{tabular}{c|c||c}
\hline\hline
$i$  & ${\cal M}^{(i)}_{\rm lat}$ (GeV)$^3$ & ${\cal M}^{\prime\ (i)}_{\rm lat}$ (GeV)$^3$ \\
\hline
1  & -0.247(62)	&-0.147(242)		\\
2  & 0.266(72) 	&-0.218(54)		\\
3  & -0.064(183)	& 0.295(59)		\\
4  & 0.444(189) 	& \textemdash		\\
5  & -0.601(146) 	&-0.601(146)		\\
6  & -1.188(287)	&-1.188(287)		\\
7  & 1.33(8)		& 1.33(8)			\\ 
8  & 4.65(14)		& 4.65(15)		\\ 
9  & -0.345(97) 	& \textemdash		\\
10 & 0.176(100)	& \textemdash 	\\
\end{tabular}
\caption{The unrenormalized matrix elements in the conventional ten-operator basis (second column) and the seven-operator chiral basis (third column). Only statistical errors are shown. \label{tab:Munrenorm} }
\end{table}

Table~\ref{tab:Munrenorm} shows the $K\to\pi\pi$ matrix elements of the unrenormalized lattice operators.  The second column contains the matrix elements of the ten operators $\{Q_i\}_{1\le i \le 10}$ written in the traditional, physical basis defined in Eqs. (2)-(5) of Ref.~\cite{Lehner:2011fz}.  The third column shows the matrix elements of the seven unrenormalized operators  in the chiral basis, $Q_j'$, where $j \in \{1, 2, 3, 5, 6, 7, 8\}$.  These are defined in Eq.~(10) of Ref.~\cite{Lehner:2011fz}.  For the convenience of the reader, we have converted the values in this table into physical units and have applied the Lellouch-L\"{u}scher finite-volume correction of $F= 23.96(61)$ to obtain infinite-volume matrix elements with the normalization conventions specified in Ref.~\cite{Blum:2012uk}.  

Only seven of the ten operators shown in the second column of Table~\ref{tab:Munrenorm} are independent; the remaining three are related to those by Eq. (9) of Ref.~\cite{Lehner:2011fz}.  As we make use of stochastic sources to approximate the high eigenmodes of our all-to-all propagators, our matrix elements only obey these relations at the percent scale.  To convert to the chiral basis we discard the matrix elements corresponding to the operators $Q_4$, $Q_9$ and $Q_{10}$. 

\begin{table}
\begin{tabular}{c|c||c}
\hline\hline
$i$  & ${\cal M}^{\prime\ (i)}_{ {\rm SMOM}}$ (GeV)$^3$ & ${\cal M}^{(i)}_{ \overline{\rm MS} }$  (GeV)$^3$\\
\hline
1  & -0.0675(1109)(128)	&-0.151(29)(36)\\
2  & -0.156(27)(30)		& 0.169(42)(41)\\
3  & 0.212(52)(40)		&-0.0492(652)(118)\\
4  & \textemdash		& 0.271(93)(65)\\
5  & -0.193(62)(37) 		&-0.191(48)(46)\\
6  & -0.366(103)(70) 	&-0.379(97)(91)\\
7  & 0.225(37)(43)		& 0.219(37)(53)\\
8  & 1.65(5)(31) 		& 1.72(6)(41)\\
9  & \textemdash		&-0.202(54)(49) \\
10 & \textemdash		& 0.118(42)(28) \\
\end{tabular}
\caption{The renormalized matrix elements in the RI/SMOM$(\slashed q,\slashed q)$ scheme and chiral basis (second column).  The third column shows the matrix elements of the traditional, physical operators $Q_i^{\overline{\mathrm{MS}}}$ defined in the $\overline{\rm MS}$ scheme. The latter are obtained by applying the $10\times 7$ scheme-change matrix factors given in Table~\ref{tab:rimomtoMSbarcoeffs} to the numbers in the second column. The left error shown is statistical and the right is systematic.  For the second column the systematic error is a uniform 19\% estimate obtained from Table II by omitting the errors associated with operator renormalization, input parameters and Wilson coefficients.  (This estimate ignores the errors associated with our use of an incomplete set of off-shell, RI/SMOM operators, an error which we perturbatively estimate at the 1\% level.)  The errors in the third column are similarly obtained from Table II except only the parametric and Wilson coefficient errors are omitted, giving a uniform 24\% systematic error estimate. \label{tab:Mrenorm}}
\end{table}

\begin{table*}
\begin{equation}
\left(
\begin{array}{ccccccc}
0.4582804(281) & 0 & 0 & 0 & 0 & 0 & 0 \\
0 & 0.2928(300) & -0.3130(349) & -0.0123(112) & 0.00579(714) & 0 & 0 \\
0 & 0.1368(481) & 0.8350(573) & -0.0051(209) & 0.0060(139) & 0 & 0 \\
0& -0.160(134) & -0.160(143) & 0.4393(486) & -0.0707(286) & 0 & 0 \\
0 & 0.0335(786) & 0.106(100) & -0.0355(367) & 0.3456(239) & 0 & 0 \\
0 & 0 & 0 & 0 & 0 & 0.4923552(284) & -0.0921608(285) \\
0 & 0 & 0 & 0 & 0 & -0.0593122(599)& 0.372432(125) 
\end{array}\right)
\nonumber
\end{equation}
\caption{The $7\times 7$, lattice-to-RI/SMOM$(\slashed q,\slashed q)$ conversion matrix $Z^{\mathrm{lat}\to\mathrm{RI}}_{kj}$.  Only statistical errors are shown. \label{tab:rimomcoeffs} }
\end{table*}

\subsection{Non-perturbative renormalization}

The seven unrenormalized operators $Q_j'$ in the chiral basis are transformed to the operators $Q_k^{\mathrm{RI}}$, renormalized in the non-perturbatively defined RI/SMOM$(\slashed q,\slashed q)$ scheme, by the $7\times 7$ matrix $Z^{\mathrm{lat}\to\mathrm{RI}}_{kj}$:
\begin{equation}
Q_k^{\mathrm{RI}} = \sum_j Z^{\mathrm{lat}\to\mathrm{RI}}_{kj}Q_j', 
\end{equation}
for $k \in \{1, 2, 3, 5, 6, 7, 8\}$.  The matrix $Z^{\mathrm{lat}\to\mathrm{RI}}_{kj}$
is determined by requiring that the $Q_k^{\mathrm{RI}}$ operators (after the subtraction of dimension-three and dimension-four operators) obey the RI/SMOM$(\slashed q,\slashed q)$ normalization conditions specified in Ref.~\cite{Lehner:2011fz}.  These are conditions imposed on amputated, Landau-gauge, four-fermion Green's functions which contain the operators $Q_k^{\mathrm{RI}}$ and are evaluated at off-shell, external momenta that are specified by the two momenta $p_1$ and $p_2$ .  In our case $p_1 = \frac{2\pi}{L}(0,4,4,0)$ and $p_2 = \frac{2\pi}{L}(4,4,0,0)$ which together satisfy the symmetric momentum condition $p_1^2 = p_2^2 = (p_1-p_2)^2 = (\mu)^2$, where $L=32$ is the spatial extent of the lattice. Here $\mu= 1.53$ GeV determines the renormalization scale.  The resulting matrix elements of these transformed RI/SMOM$(\slashed q,\slashed q)$ operators are given in Table~\ref{tab:Mrenorm} and the NPR matrix  $Z^{\mathrm{lat}\to\mathrm{RI}}_{kj}$ in Table~\ref{tab:rimomcoeffs}.

Finally we use the one-loop perturbative formulae given in Eqs.~(54), (56), (60) and Table XI of Ref.~\cite{Lehner:2011fz} to evaluate the matrix $Z^{\mathrm{RI}\to\overline{\mathrm{MS}}}_{ik}$ which determines the ten traditional $\overline{\mathrm{MS}}$ operators $Q^{\overline{\mathrm{MS}}}_i$ in terms of the seven chiral basis operators $Q^{\mathrm{RI}}_j$:
\begin{equation}
Q^{\overline{\mathrm{MS}}}_i = \sum_k Z^{\mathrm{RI}\to\overline{\mathrm{MS}}}_{ik}Q^{RI}_k.
\end{equation}
It is the matrix elements of these ten $\overline{\mathrm{MS}}$ operators, multiplied by the appropriate Wilson coefficients, which are given in Table~\ref{tab:A0_by_op}.

\begin{table*}
\begin{equation}
\left(
\begin{array}{ccccccc}
0.19744	& 1.07389	& 0.13830	& 0			& 0			& 0			& 0 \\
0.19744	&-0.12667	& 0.75559	& 0.00208	&-0.00625	& 0			& 0 \\
0		& 2.96832	& 1.92607	& 0.00417	&-0.01250	& 0			& 0 \\
0 		& 1.74719	& 2.49297	& 0.01554 	&-0.04636	& 0			& 0 \\
0 		& 0 			& 0 			& 1.00121	&-0.00364	& 0			& 0 \\
0 		&-0.04687 	& -0.10936	&-0.00164	& 0.99519	& 0			& 0 \\
0 		& 0 			& 0 			& 0			& 0 			& 1.00121	&-0.00364 \\
0 		& 0 			& 0 			& 0			& 0 			&-0.01726 	& 1.04206 \\
0.29616	& 0.12667	& -0.75559 	& -0.00208	& 0.00625 	& 0 			& 0 \\
0.29616	&-1.07389	& -0.13830 	& 0			& 0 			& 0			& 0 \\
\end{array}
\right)
\nonumber
\end{equation}

\caption{The $10\times 7$ RI/SMOM$(\slashed q,\slashed q)\to \overline{\rm MS}$ conversion matrix, $Z^{\mathrm{RI}\to\overline{\mathrm{MS}}}_{ik}$ computed to one-loop in QCD perturbation theory at the scale $\mu=1.531$ GeV. \label{tab:rimomtoMSbarcoeffs}}

\end{table*}

\begin{table}[t]
\begin{tabular}{ccc}
\hline\hline
$i$ & $z_i$ & $y_i$ \\
\hline
1 & -0.3734				&  0 			\\
2 & 1.189					&  0			\\
3 & 0.001294				&  0.02720	\\
4 & -0.003691				& -0.05828 	\\ 
5 & 0.002505				&  0.007133	\\ 
6 & -0.004718 				& -0.08191	\\ 
7 & 6.099$\times 10^{-5}$	& -0.0003401	\\
8 & -4.414$\times 10^{-5}$	&  0.0008684	\\ 
9 & 3.561$\times 10^{-5}$	& -0.01046	\\
10 & 3.010$\times 10^{-5}$	&  0.003496
\end{tabular}
\caption{The Wilson coefficients in the $\overline{\rm MS}$ scheme at $\mu=1.531$ GeV. \label{tab:wilsoncoeffs} }
\end{table}

\subsection{Wilson coefficients}

\begin{table*}
\begin{tabular}{c|c}
\hline\hline
Input & Value  \\
\hline
$\mu$ & 1.531 GeV \\
$m_W$ & 80.385 GeV \\
$m_t$ & 160.0 GeV \\
$m_c$ & 1.275 GeV \\
$m_b$ & 4.18 GeV \\
$\alpha_{\rm EM}^{-1}$ & 127.940 \\ 
$\sin^2(\theta_W)$ & 0.23126 \\
$\alpha_s(m_Z)$ & 0.1185 \\
$\Lambda_4$ & 0.331416 GeV\\
\hline
$\tau$ & $0.001543 -0.000635i$  \\
$V_{us}$ & 0.2253 \\
$V_{ud}$ & 0.97425 \\
$\omega$ & 0.04454\\
$|\varepsilon|$ & 0.002228 \\
$\phi_\varepsilon$ & 0.75957 rad\\
${\rm Re}(A_0)_{\rm expt}$ & 3.3201$\times 10^{-7}$ GeV\\
${\rm Re}(A_2)_{\rm expt}$ & 1.479$\times 10^{-8}$ GeV\\
\end{tabular}
\caption{Inputs for the computation of the Wilson coefficients are given above the break, and the remaining inputs that were used for computing ${\rm Re}(\varepsilon'/\varepsilon)$ are given below.  The parameters given in this table were obtained from the PDG Review of Particle Physics~\cite{Agashe:2014kda}. Here $\phi_\varepsilon$ is the phase of the indirect CP-violation parameter $\varepsilon$. The last two entries are the experimental values of Re($A_0$) and Re($A_2$), which we combine with our lattice values for the imaginary components to obtain ${\rm Re}(\varepsilon'/\varepsilon)$. Other than the systematic errors listed in Table~\ref{tab:sys_errors}
we neglect the errors on these inputs as their contributions to our final error are small in relation to our statistical errors.
\label{tab:inputconsts} }
\end{table*}

The real and imaginary parts of the amplitude $A_0 = \langle (\pi\pi)_{I=0}|H_W|K\rangle$ are obtained by multiplying the $\overline{\rm MS}$-renormalized matrix elements by the appropriate Wilson coefficients and CKM matrix elements as given in Eq.~\eqref{eq:H_W}.
 
The Wilson coefficients were computed at the renormalization scale of $\mu=1.531$ GeV from the equations given in Ref.~\cite{Buchalla:1995vs}.  For the required standard model parameters we use the values listed in Table~\ref{tab:inputconsts}, from which we obtain the Wilson coefficients shown in Table~\ref{tab:wilsoncoeffs}. We use the two-loop beta function to obtain the three-flavor value of $\alpha_S = 0.353388$ at $\mu = 1.531$ GeV. The CKM matrix elements $V_{us}$ and $V_{ud}$, along with the remaining inputs necessary to compute ${\rm Re}(\varepsilon'/\varepsilon)$ from Eq.~\eqref{eq:ep-e_theory} are also listed in Table~\ref{tab:inputconsts}.

\end{document}